\documentclass[aps, prd, reprint, twocolumn, superscriptaddress]{revtex4-1}
\usepackage{amsmath}
\usepackage{amssymb}
\usepackage{graphicx}
\usepackage{color}
\usepackage{times}
\usepackage{euscript}
\usepackage{amsthm}
\usepackage{amsfonts}
\usepackage[english]{babel}
\usepackage{epstopdf}
\usepackage{multirow,makecell,array}
\usepackage{mathtools}
\usepackage{float}
\usepackage[dvipsnames]{xcolor}

\begin{document}

\title{Vacuum pair production in zeptosecond pulses: Peculiar momentum spectra \\ and striking particle acceleration by bipolar pulses}

\author{I.~A.~Aleksandrov}
\affiliation{Department of Physics, Saint Petersburg State University, Universitetskaya Naberezhnaya 7/9, Saint Petersburg 199034, Russia}
\affiliation{Ioffe Institute, Politekhnicheskaya Street 26, Saint Petersburg 194021, Russia}
\author{N.~N. Rosanov}
\affiliation{Ioffe Institute, Politekhnicheskaya Street 26, Saint Petersburg 194021, Russia}

\begin{abstract}
We examine the phenomenon of electron-positron pair production from vacuum in a combination of two counterpropagating electromagnetic pulses having a duration of the order of the Compton time. We show that in this extreme short-time domain, the momentum distributions of the particles produced possess a peculiar structure which strongly depends on whether the electromagnetic pulses have a unipolar or bipolar profile. It is shown that bipolar pulses can predominantly generate particles with ultrarelativistic velocities along the propagation direction of the pulses, while unipolar ones are generally more favorable in terms of the total particle yield in the same regime. The highly nontrivial properties of the $e^+e^-$ spectra revealed in our study provide strong experimental signatures paving the way to probe a complex vacuum response within the short-time domain of quantum electrodynamics.
\end{abstract}

\maketitle

{\it Introduction.} Almost a century ago, in the 1930s, a series of theoretical investigations unveiled remarkable properties of the quantum vacuum encompassing continual events of production and annihilation of virtual electron-positron pairs and their polarizability. It was demonstrated that the quantum nature of the vacuum underlies its nontrivial response to external perturbations and gives rise to fascinating nonlinear phenomena such as light-by-light scattering~\cite{euler_kockel, weisskopf, heisenberg_euler, karplus_pr_1950, karplus_pr_1951} and Sauter-Schwinger $e^+e^-$ pair production~\cite{sauter_1931, heisenberg_euler, schwinger_1951} (for review, see, e.g., Refs.~\cite{dittrich_gies, dunne_shifman, dipiazza_rmp_2012, king_heinzl_2016, blaschke_review_2016, xie_review_2017, fedotov_review}). Probing the properties of the quantum vacuum has been a strong and exciting challenge since then and has involved huge theoretical and experimental efforts.

As the nonlinear vacuum effects yield larger observable signals in the presence of extremely strong external fields, one strives to take advantage of the rapid and continuing advancement of laser technologies, which makes ever higher intensities available for practical investigations~\cite{fedotov_review}. Although the total energy carried by extreme laser pulses does not exceed thousands of Joules, it is their tremendously short duration that gives rise to huge field amplitudes. In this paper, we focus on the regime of ultrashort pulses having a duration of the order of the Compton time $\tau_\text{C} \sim 1~\text{zs} = 10^{-21}~\text{s}$. Nowadays, one witnesses a significant progress in the generation of (sub)attosecond pulses (see, e.g., Refs.~\cite{krausz_rmp_2009, popmintchev_nat_ph_2010, ipp_plb_2011, midorikawa_2011, takahashi_2013, li_nat_comm_2020, shim_2022, witting_2022} and references therein), and there are already many proposals concerning the zeptosecond regime~\cite{gordienko_prl_2004, klaiber_arxiv_2007, mourou_science_2011, andreev_2011, wang_prr_2022}. Further exploring the field of ultrafast physics, we examine the fundamental phenomenon of vacuum $e^+e^-$ pair production in the regime of extremely short interactions.

The effect of ultrashort electromagnetic pulses on ``ordinary'' quantum objects, such as an electron, atom, molecule, or nanoparticle, is mainly governed by the electric-field area of the pulse,
\begin{equation}
\mathbf{S}_E = \int \limits_{-\infty}^{+\infty} \! \mathbf{E}(t) dt, \label{eq:S_E_def}
\end{equation}
where the electric field strength $\mathbf{E}$ is integrated over time $t$ (see, e.g., reviews~\cite{arkhipov_qe_2020,rosanov_2024}, Refs.~\cite{dimitrovski_2005,leblond_2008,song_2010,kozlov_pra_2011,leblond_2013,glazov_pra_2024}, and references therein). The greatest impact on a given quantum system is exerted by {\it unipolar} pulses due to their unidirectional action on electric charges (for recent advancements, see, e.g., Refs.~\cite{moskalenko_2017,arkhipov_2020,rosanov_pra_2021}). One would expect that unipolar pulses would lead to more efficient production of pairs from the vacuum, ``pulling'' the electrons and positrons in opposite directions. However, as will be seen below, within the domain of short-time quantum electrodynamics (QED), our quasiclassical intuitive perspectives are no longer valid. In particular, we will demonstrate that a ``collision'' of two linearly polarized pulses of a zeptosecond duration can make the vacuum generate ultrarelativistic pairs and push them perpendicularly to the electric field component. As will be shown, the properties of the $e^+e^-$ momentum distributions are extremely sensitive to the phase parameter governing the pulse profiles. Throughout the paper, we employ the units $\hbar = c = 1$, $\alpha = e^2/(4\pi) \approx 1/137$. The electron mass and charge are denoted by $m$ and $e$, respectively. The Compton time then reads $\tau_\text{C} = m^{-1} \approx 1.3~\text{zs}$, and the Schwinger critical strength is $E_\text{cr} = m^2/|e| \approx 1.3 \times 10^{16}~\text{V}/\text{cm}$.

\mbox{}

\begin{figure}[t]
\center{\includegraphics[width=0.95\linewidth]{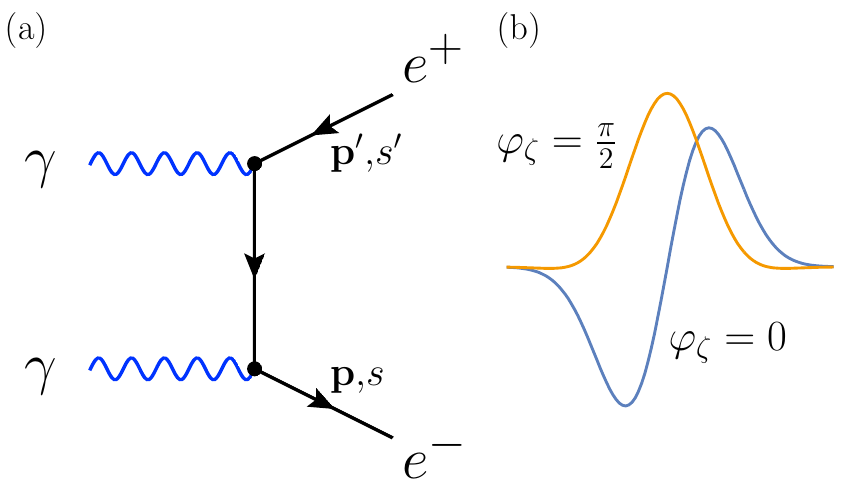}}
\caption{(a) Leading-order Feynman diagram describing pair production. In our setup, the photon lines correspond to two counterpropagating pulses. (b) Electric field profile~\eqref{eq:e_zeta} for $\sigma = 1$ and two different values of the phase $\varphi_\zeta$ corresponding to a unipolar pulse and bipolar pulse, respectively. The latter was rescaled, so that the pulse energy density~\eqref{eq:W} is the same as in the unipolar setup.}
\label{fig:BW}
\end{figure}

{\it Setup.} We consider a combination of two linearly polarized counterpropagating pulses having amplitude $E_0$ and duration $\tau$. Since in the regime of extremely small pulse durations ($\tau \sim \tau_\text{C}$), this setup cannot obviously be treated as a classical external background, we should instead consider the Breit-Wheeler process~\cite{BW}, where two photons convert into an electron-positron pair. The corresponding Feynman diagram within the leading order of perturbation theory (PT) is depicted in Fig.~\ref{fig:BW}(a). Although the (non)linear Breit-Wheeler mechanism involving monochromatic initial photons or long-duration pulses has been extensively discussed in numerous investigations (see, e.g., Refs.~\cite{reiss_1962,nikishov_ritus_1964,ivanov_2005,heinzl_plb_2010,titov_prl_2012,krajewska_pra_2012,titov_pra_2013,di_piazza_prl_2016,seipt_pra_2020,golub_prd_2021,tang_prd_2021,podszus_prd_2022}), here we will explore a completely different QED regime, where the quantum vacuum interacts with two extremely short perturbations. The essential point here is that the actual expression for the Feynman diagram in Fig.~\ref{fig:BW}(a) has a universal form that does not distinguish between a classical background and photons. In other words, employing ``photon wave functions'' with a short Gaussian envelope or using a classical electromagnetic pulse, one will obtain identical results. The situation becomes a bit more involved in the case of unipolar pulses, for which the number of zero-energy quanta is formally infinite. In this case, one can represent the pulse as a sum of a high-frequency part consisting of photons and a low-frequency part which should be treated as a classical background. Nevertheless, if the pair-production process can be accurately described by means of the PT expansion with respect to both the quantum and classical components, one can combine them straight away since the leading-order PT terms are equivalent as was pointed out above. To justify our approach, it is therefore sufficient to make sure that the higher-order contributions (diagrams with more vertices) are small, so the leading-order term covers the whole effect. The validity of PT is achieved due to the extremely short pulse duration $\tau$ examined in the present study.

Although the higher-order contributions can be neglected, it is vital to take into account the {\it spatial} dependence of the pulse profile, so that its quanta possess a correct energy-momentum relation and thus properly reproduce the kinematics of the process. The two electromagnetic pulses will be described by the following $x$-projection of the electric-field component:
\begin{eqnarray}
E^{(\zeta)}_x (t,z) &=& E_0 e^{(\zeta)} (\sigma t/\tau - \zeta \sigma z/\tau), \label{eq:external_field} \\
e^{(\zeta)} (\eta) &=& \mathrm{e}^{-(\eta/\sigma)^2} \sin (\eta + \varphi_\zeta),
\label{eq:e_zeta}
\end{eqnarray}
where $\zeta = +$ corresponds to the pulse traveling along the $z$ axis, while the pulse with $\zeta  =-$ propagates in the opposite direction. Here $\sigma$ indicates the number of periods of the sine in Eq.~\eqref{eq:e_zeta} within the pulse envelope. The phase parameters $\varphi_\zeta$ determine the electric-field area~\eqref{eq:S_E_def} and govern a transition from unipolar ($\varphi_\zeta = \pi/2$) to bipolar ($\varphi_\zeta = 0$) pulses [see Fig.~\ref{fig:BW}(b)]. The electric-field area reads
\begin{equation}
S^{(\zeta)}_E = \sqrt{\pi} E_0 \tau \mathrm{e}^{-\sigma^2/4} \sin \varphi_\zeta .
\label{eq:S_E_explicit}
\end{equation}
The small parameter within the PT expansion, is generally the classical nonlinearity parameter $\xi = (E_0/E_\text{cr}) (\tau / \tau_\text{C})$~\cite{brezin}, or, equivalently, in the case of unipolar pulses, the ratio $|eS_E|/m$. Given a very small duration $\tau$, the validity region of PT is very broad in terms of the amplitude $E_0$. Let us also introduce the integral of $E^2(t)$ over $t$ representing the energy density of the pulse:
\begin{equation}
W^{(\zeta)} = \frac{\sqrt{\pi}}{2^{3/2}} \, E_0^2 \tau \big ( 1 - \mathrm{e}^{-\sigma^2/2} \cos 2\varphi_\zeta \big ).
\label{eq:W}
\end{equation}

In fact, as we are interested in ultrashort pulses including unipolar ones, we generally consider $\sigma \sim 1$, so the concepts of the pulse carrier and its frequency are not well defined. In this regime, the value of $\varphi_\zeta$ significantly affects the shape of the pulse and its electric-field area. We will fix $\sigma = 1$ and vary $\varphi_\zeta$. In Fig.~\ref{fig:BW}(b) we depict an individual pulse for two different values of the corresponding phase $\varphi_\zeta$. For $\varphi_\zeta = \pi/2$ the field profile takes almost everywhere non-negative values. For $\varphi_\zeta = 0$ it has a zero electric-field area. In this {\it bipolar} case, we multiply the field profile by factor $(\tanh \sigma^2/4)^{-1/2}$, so that the both fields have the same value of $W^{(\zeta)}$ given in Eq.~\eqref{eq:W}. In what follows, we will demonstrate that the pair-production process exhibits surprisingly different patterns in the unipolar scenario ($\varphi_\zeta = \pi / 2$) and bipolar one ($\varphi_\zeta = 0$).

\mbox{}

{\it Theoretical description.} To provide a quantitative and qualitative description of the pair-production process, we will compute the momentum distributions of the particles created. We evaluate the Feynman diagram presented in Fig.~\ref{fig:BW}(a). The only nonzero contribution arises when the external photon lines correspond to {\it different} electromagnetic pulses. Our goal is to evaluate the electron number density $f (\mathbf{p})$ as a function of the particle's momentum~$\mathbf{p}$. As our setup is infinite in the transverse $xy$ plane, the function $f (\mathbf{p})$ will be defined via $f (\mathbf{p}) = [(2\pi)^2/S] n(\mathbf{p})$, where $S$ is the $xy$ cross-section area and $n(\mathbf{p})$ is the mean value of the number-density operator within the second-quantization approach. The {\it total} number of pairs per unit normalization area is then calculated via
\begin{equation}
\mathcal{N} = \int \! f(\mathbf{p}) \, d\mathbf{p}.
\label{eq:N}
\end{equation}
The electron number density reads (see Supplemental Material~\cite{SM})
\begin{eqnarray}
f (\mathbf{p}) &=& \frac{1}{\pi^2} \bigg ( \frac{E_0}{E_\text{cr}} \bigg )^4 \bigg ( \frac{m\tau}{\sigma} \bigg )^4
\! \int \! dp_z' \, G(\mathbf{p}, p_z') \nonumber \\
{} &\times& \big | F^{(+)} (q_+) F^{(-)} (q_-) \big |^2,
\label{eq:f_PT}
\end{eqnarray}
where $q_\pm = [p_0 + p_0' \pm (p_z + p_z')]\tau/(2\sigma)$, $p_0 = \sqrt{m^2 + \mathbf{p}^2}$, $p_0' = \sqrt{m^2 + p_x^2 + p_y^2 +p_z'^2}$, and
\begin{multline}
F^{(\zeta)} (q) = \sqrt{\pi} \sigma \mathrm{e}^{-\sigma^2 (q^2+1)/4} \\
{}\times \bigg [ \sin \varphi_\zeta \cosh \bigg ( \frac{\sigma^2 q}{2} \bigg ) + i \cos \varphi_\zeta \sinh \bigg ( \frac{\sigma^2 q}{2} \bigg ) \bigg ]
\label{eq:F}
\end{multline}
is the Fourier transforms of the profiles~\eqref{eq:e_zeta} ($\zeta = \pm$). The function $G(\mathbf{p}, p_z')$ is independent of the setup parameters (its explicit form is given in Supplemental Material~\cite{SM}). As will be shown below, here it is crucial whether the external-field profile is unipolar ($\varphi_\zeta = \pi / 2$) or bipolar ($\varphi_\zeta = 0$).

\mbox{}

\begin{figure}[t]
\center{\includegraphics[width=0.98\linewidth]{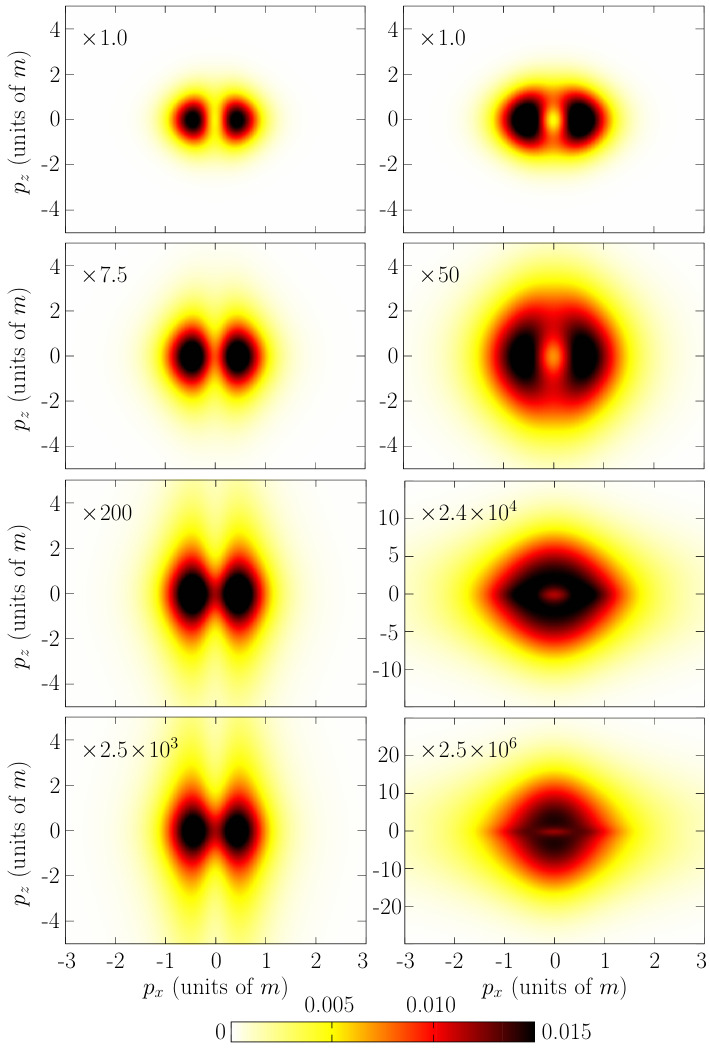}}
\caption{Momentum distributions~\eqref{eq:f_PT} of the electrons produced for $p_y = 0$ and $\tau / \tau_\text{C} = 1.0$, $0.5$, $0.2$, and $0.1$ (from top to bottom). The left column corresponds to the combination of unipolar pulses ($\varphi_+ = \varphi_- = \pi/2$), while the right column was obtained for the bipolar setup ($\varphi_+ = \varphi_- = 0$). The density~\eqref{eq:f_PT} was multiplied by the numerical factors indicated in the graphs.}
\label{fig:spectra}
\end{figure}

{\it Numerical results.} We fix $\sigma = 1$ and perform the calculations of the electron number density $f(\mathbf{p})$ for various $\tau$. The amplitude is formally chosen as $E_0 = E_\text{cr}$ for unipolar pulses and additionally adjusted for bipolar pulses to preserve the same value of $W^{(\zeta)}$: in the case of bipolar pulses, we always choose $E_0 = (\tanh \sigma^2/4)^{-1/2} E_\text{cr} = 2.02 E_\text{cr}$. For other values of $E_0$ the results can be easily rescaled according to the prefactor $(E_0/E_\text{cr})^4$. To obtain the accurate predictions, one has to make sure that $\xi \ll 1$. In Fig.~\ref{fig:spectra} we present the $p_x p_z$ momentum distributions for $\varphi_+ = \varphi_- = 0$ and $\varphi_+ = \varphi_- = \pi/2$. We observe that the number densities strongly depend on the phases $\varphi_\zeta$ and for $\tau < \tau_\text{C}$ provide much larger values in the case of unipolar pulses ($\varphi_+ = \varphi_- = \pi/2$).

This can be understood by inspecting the integrand in Eq.~\eqref{eq:f_PT}. For $\mathbf{p} = 0$, the only energy scale in the $G$ function is the electron mass $m$, so this function decreases for $|p_z'| \gtrsim m$. On the other hand, the Fourier transforms for $|p_z'| \lesssim m$ are mainly governed by either the $\sinh$ or $\cosh$ in Eq.~\eqref{eq:F}. In the latter (unipolar) case, the overlap between $G$ and the Fourier transforms is significantly larger, which yields much larger values of the number density. This indicates that ultrashort unipolar pulses are substantially more efficient in terms of pair generation. Although the full result~\eqref{eq:f_PT} represents a rather nontrivial expression, in the limiting case $\tau \ll \tau_\text{C}$, it is possible to obtain simple analytical estimates which show that in the bipolar case the particle density is suppressed by the additional factor $(\tau/\tau_\text{C})^3$ compared to the unipolar-field results (see Supplemental Material~\cite{SM}). 

Another strong signature of the electron spectra is related to its support with respect to $p_z$. In Fig.~\ref{fig:spectra} we clearly see that whereas in the unipolar case $\varphi_+ = \varphi_- = \pi/2$ the momentum distributions vanish for $|p_z| \gtrsim m$, bipolar pulses ($\varphi_+ = \varphi_- = 0$) produce pairs with extremely large energies once $\tau \lesssim \tau_\text{C}$ (see the right column in Fig.~\ref{fig:spectra}). From the computational perspective, this remarkable feature can be again explained by considering the integrand in Eq.~\eqref{eq:f_PT}. Although the number density itself is significantly smaller in this case, the relative contributions governed by the $\sinh$ function in Eq.~\eqref{eq:F} are always large as long as $|p_z| \lesssim 1/\tau$. It turns out that in the case of extremely short pulses, the bipolar setup creates unltrarelativistic particles mainly propagating along the $z$ direction. Since the electric field component exert a force in the $x$ direction, we find here a very nontrivial feature which may seem somewhat counterintuitive. It is the structure of the Fourier transform of the external field that strongly depends on $\varphi_\zeta$ and becomes responsible for the peculiar properties of the momentum distributions and, in particular, for the resulting momentum transfer. Finally, we note that the $p_y$ spectrum was found to be a regular bell-shaped profile of the width $\sim m$ in all of our calculations.

\begin{figure}[b]
\center{\includegraphics[width=0.98\linewidth]{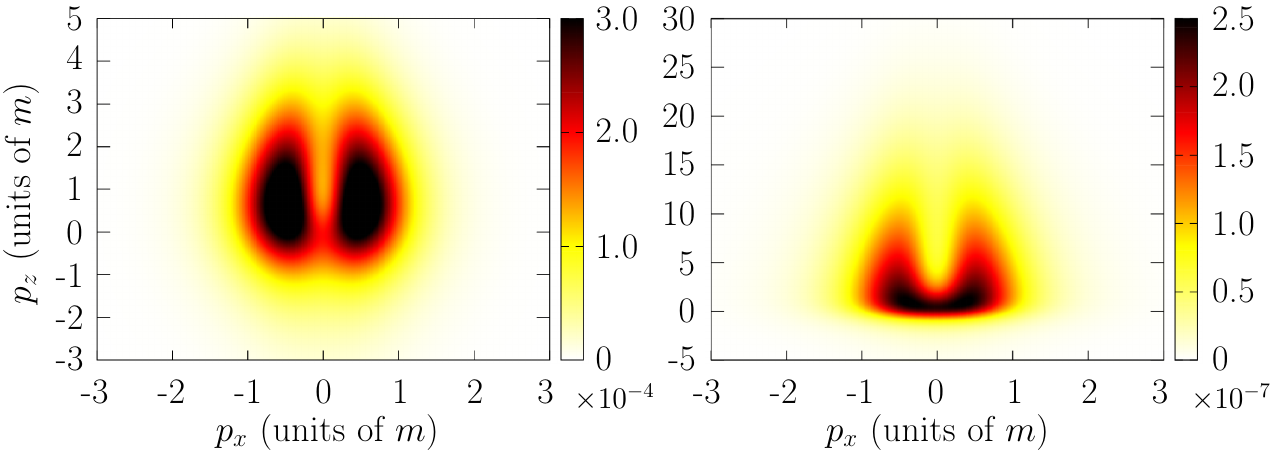}}
\caption{Momentum distributions~\eqref{eq:f_PT} of the electrons produced with $p_y = 0$ for $\tau = 0.5 \tau_\text{C}$ (left) and $\tau = 0.1 \tau_\text{C}$ (right). The phase parameters are $\varphi_+ = 0$ and $\varphi_- = \pi/2$ corresponding to a combination of a unipolar pulse and bipolar one.}
\label{fig:spectra_phi}
\end{figure}

In Fig.~\ref{fig:spectra_phi} we display the electron spectra in the case $\varphi_+ = 0$, $\varphi_- = \pi/2$ corresponding to the combination of a unipolar field and bipolar pulse. Here the spectra are no longer symmetric with respect to $p_z \to -p_z$ which represents one more distinctive measurable signature of the momentum distributions.

\begin{figure}[t]
\center{\includegraphics[width=0.97\linewidth]{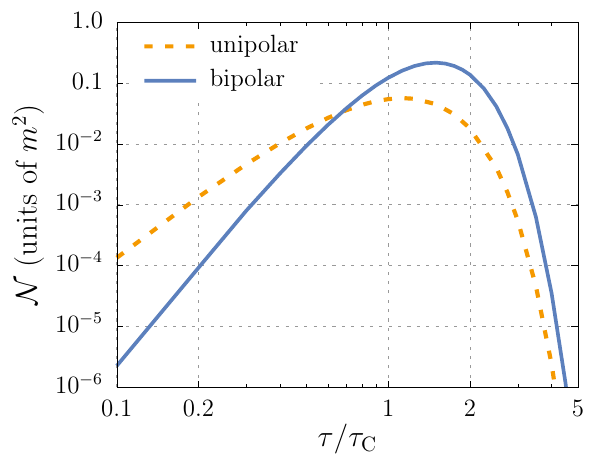}}
\caption{Total number of pairs produced in the case of the unipolar setup (dashed line) and bipolar pulses (solid line) as a function of the pulse duration $\tau$. The amplitude of the bipolar pulses is rescaled, so that the energy density~\eqref{eq:W} is the same in the two scenarios.}
\label{fig:N}
\end{figure}

As was uncovered above, in the regime $\tau < \tau_\text{C}$ unipolar pulses are more efficient in terms of providing larger {\it differential} particle numbers $f (\mathbf{p})$. On the other hand, for $\tau = \tau_\text{C}$ the number densities presented in the first row in Fig.~\ref{fig:spectra} are of the same level. Moreover, according to our results obtained for $\tau > \tau_\text{C}$, the momentum distributions are qualitatively the same as for $\tau = \tau_\text{C}$, but the number densities become significantly smaller. To clarify these aspects, let us now evaluate the {\it total} pair yield~\eqref{eq:N} by integrating $f(\mathbf{p})$ over three-dimensional momentum $\mathbf{p}$. The results are presented in Fig.~\ref{fig:N} as a function of $\tau$. It is now evident that in the regime $\tau \lesssim 0.5 \tau_\text{C}$ the total particle yield in the case of unipolar pulses can be substantially larger than that obtained for bipolar pulses. As was pointed out above, the number densities in the case of bipolar pulses are smaller, and for $\tau \ll \tau_\text{C}$, they are suppressed by the factor $(\tau/\tau_\text{C})^3$. On the other hand, the particle spectra are considerably broader in the case of the bipolar setup. Since the corresponding width with respect to $p_z$ scales as $1/\tau$, the total pair yield is suppressed by $(\tau/\tau_\text{C})^2$. These findings clearly indicate that unipolar pulses are indeed more favorable for enhancing the pair-production effect in the domain $\tau \lesssim 0.5 \tau_\text{C}$. Nevertheless, Fig.~\ref{fig:N} also demonstrates that for larger values of $\tau$, bipolar pulses become more productive. In this domain, our qualitative analysis of Eq.~\eqref{eq:f_PT} should be different as $q_\pm$ contain now a large factor $\tau$. As $p_0 \geqslant m$, at least one of the quantities $q_+$ and $q_-$ is now always large and thus yields a very small factor coming from the Fourier transform~\eqref{eq:F}. For $\tau \gg \tau_\text{C}$, the number of pairs becomes many orders of magnitude smaller (cf.~Refs.~\cite{titov_prl_2012,titov_pra_2013}), but within this region, one has to carefully assess the validity of perturbation theory. As clearly seen from Fig.~\ref{fig:N}, most efficient pair production occurs at $\tau \sim \tau_\text{C}$ although we can hardly judge in advance which of the two field configurations is more promising. We also point out that the maximum particle yield around $\tau \sim \tau_\text{C}$ can qualitatively predicted by the analysis of simplified purely time-dependent setups~\cite{ren_cpl_2012, kohlfuerst_prd_2013}.

We underline that one could hardly expect the above patterns in the Sauter-Schwinger process involving quasistatic strong external fields, and it is a short-time domain where these striking features of the momentum spectra can arise. In the nonperturbative regime, the pair-production probability is generally suppressed by the factor $\mathrm{exp} [-\pi (m^2 + \mathbf{p}_\perp^2)/|eE_0| ]$~\cite{nikishov_constant}, which does not allow the particles to have large transverse momenta, i.e., it requires $|\mathbf{p}_\perp| \lesssim m$. Although the basic properties of the Sauter-Schwinger mechanism in couterpropagating pulses (see, e.g., Refs.~\cite{nikishov_constant,ruf_prl_2009,hebenstreit_prl_2009,bulanov_prl_2010,hebenstreit_prl_2011,dumlu_prd_2011, dumlu_prd_2011_2, ren_cpl_2012, kohlfuerst_prd_2013,woellert_prd_2015,aleksandrov_2017_2,ababekri_prd_2019,aleksandrov_kohlfuerst,kohlfuerst_prl_2022,kohlfuerst_arxiv_2022_2,li_pra_2023,hu_prd_2023,tkachev,aleksandrov_kudlis}) are well elaborated and intuitively comprehensible, our findings obtained within the short-time regime do not seem that natural since the process here cannot be understood from the quasiclassical perspective (note also that the interference effects strongly depending on the pulse shape within a strong-field domain~\cite{dumlu_prd_2011,dumlu_prd_2011_2} are not that pronounced in the spectra revealed above). Our study demonstrates that the {\it short-time} QED represents a completely different realm which urgently requires experimental verification.

With regard to the experimental prospects, we note that the zeptosecond pulses in our setup should not only consist of MeV photons, but also contain a sufficiently large amount of them. In terms of our parameters, the latter condition concerns the amplitude $E_0$. First, according to Fig.~\ref{fig:N}, for $\tau \sim \tau_\text{C}$ and $E_0 = E_\text{cr}$ the mean total pair number $N = \mathcal{N} S/(2\pi)^2$ is of the order of 0.01 (we assume $S \sim \tau_\text{C}^2$). Second, $E_0 = E_\text{cr}$ yields $\sim 100~\text{MW}$ in terms of the peak radiation power. This value is quite realistic in the domain of attosecond pulses (see, e.g., Refs.~\cite{takahashi_2013, li_nat_comm_2020, shim_2022} and references therein). Moreover, it is also possible to reach a GW~\cite{takahashi_2013} or even TW regime~\cite{shim_2022}. Since the peak power is proportional to $E_0^2$ and $N \sim E_0^4$, one can easily rescale our results to estimate the pair signal for a given setup, provided the PT approach remains justified. Our numerical predictions concern a {\it single} ``collision'' of two ultrashort pulses, so it is also necessary to take into account the number of these events under specific experimental conditions. Here we note that a repetition rate of $1~\text{kHz}$ is a standard value for modern attosecond laser sources (see, e.g., Refs.~\cite{li_nat_comm_2020, witting_2022}). The above estimates demonstrate how the total particle yield depends on the parameters of zeptosecond pulses to be generated with future facilities.

Finally, we point out that the pair-production process in ultrashort fields can effectively occur in relativistic heavy-ion collision (see, e.g., Refs.~\cite{baur_sr_2007,najjari_pra_2009} and references therein), but in the setup considered in our study one can directly probe the effects regarding the unipolarity degree of the external field giving rise to the remarkable properties of the momentum distributions revealed above.

\mbox{}

{\it Conclusion.} We investigated the phenomenon of electron-positron pair production from the vacuum state. Pursuing the extreme of ultrashort fields, we demonstrated that a combination of two counterpropagating electromagnetic pulses of a zeptosecond duration can produce $e^+e^-$ pairs with peculiar momentum distributions. The properties of the particle spectra strongly depend on the pulse structure. Although in the case of extremely short pulses, it is more advantageous to generate unidirectional electric fields, a bipolar setup becomes more favorable for a larger pulse duration. Furthermore, it was shown that whereas unipolar pulses produce pairs with a well-localized momentum spectrum, bipolar ones can transfer a substantial fraction of energy making the particles ultrarelativistic. These findings provide unique signatures for probing QED in the context of ultrashort interactions.

\mbox{}

{\it Acknowledgments.} The study was funded by the Russian Science Foundation, grant No.~23-12-00012. We are also grateful to Prof.~Vladimir Kocharovsky for stimulating discussions.


\end{document}